\documentclass{iopjournal}
\pdfoutput=1

\usepackage{multirow}
\usepackage{booktabs}
\usepackage{graphicx}
\usepackage{subcaption} % 如果需要小标题
\makeatletter
\renewcommand{\articletype}[1]{\vspace*{-8mm}\noindent\Large\sf Preprint\par\vspace*{8mm}}
\renewcommand{\headrulewidth}{0pt}
\renewcommand{\footrulewidth}{0pt}
\makeatother
\hypersetup{hidelinks}

\usepackage[
  backend=bibtex,
  style=authoryear,
  maxcitenames=1,   % 1 个作者后即用 et al.
  mincitenames=1,   % 同上，确保单一作者名显示
  maxbibnames=99,   % 参考文献表可保留所有作者
  uniquename=false  % 不用首字母消歧，避免多余缩写
]{biblatex}
\DeclareFieldFormat[article]{title}{#1}
\begin{document}

\articletype{Research Paper}
\title{Neural Network-Driven Direct CBCT-Based Dose Calculation for Head-and-Neck Proton Treatment Planning}

\author{Muheng Li$^{1,2}$, Evangelia Choulilitsa$^{1,2}$, Lisa Fankhauser$^{1,2}$, Francesca Albertini$^1$, Antony Lomax$^{1,2}$ and Ye Zhang$^{1,*}$}

\affil{$^1$Center for Proton Therapy, Paul Scherrer Institute (PSI), Villigen, Switzerland}
\affil{$^2$Department of Physics, ETH Zürich, Zürich, Switzerland}
\affil{$^*$Author to whom any correspondence should be addressed.}

\email{ye.zhang@psi.ch}


% ====== PMB structured abstract ======
\begin{abstract}
\textbf{Objective.} Accurate dose calculation on cone beam computed tomography (CBCT) images is essential for modern proton treatment planning workflows, particularly when accounting for inter-fractional anatomical changes in adaptive treatment scenarios. Traditional CBCT-based dose calculation suffers from image quality limitations, requiring complex correction workflows. This study develops and validates a deep learning approach for direct proton dose calculation from CBCT images using extended Long Short-Term Memory (xLSTM) neural networks.

\textbf{Approach.} A retrospective dataset of 40 head-and-neck cancer patients with paired planning CT and treatment CBCT images was used to train an xLSTM-based neural network (CBCT-NN). The architecture incorporates energy token encoding and beam's-eye-view sequence modelling to capture spatial dependencies in proton dose deposition patterns. Training utilized 82,500 paired beam configurations with Monte Carlo-generated ground truth doses. Validation was performed on 5 independent patients using gamma analysis, mean percentage dose error assessment, and dose-volume histogram comparison.

\textbf{Main results.} The CBCT-NN achieved gamma pass rates of 95.1 ± 2.7\% using 2mm/2\% criteria. Mean percentage dose errors were 2.6 ± 1.4\% in high-dose regions ($>$90\% of max dose) and 5.9 ± 1.9\% globally. Dose-volume histogram analysis showed excellent preservation of target coverage metrics (Clinical Target Volume V95\% difference: -0.6 ± 1.1\%) and organ-at-risk constraints (parotid mean dose difference: -0.5 ± 1.5\%). Computation time is under 3 minutes without sacrificing Monte Carlo-level accuracy.

\textbf{Significance.} This study demonstrates the proof-of-principle of direct CBCT-based proton dose calculation using xLSTM neural networks. The approach eliminates traditional correction workflows while achieving comparable accuracy and computational efficiency suitable for adaptive protocols. 
\end{abstract}

\section{Introduction}
Proton therapy has established itself as a highly effective treatment modality for cancer patients. However, the unique physical properties of proton beams, characterized by the Bragg peak phenomenon, offer superior dose conformity compared to conventional photon therapy, whilst introducing heightened sensitivity to anatomical variations \parencite{newhauser2015physics}. Daily anatomical changes, including tumor shrinkage, weight loss, organ motion, and setup variations, can result in substantial alterations to dose distributions that may compromise target coverage or exceed normal tissue tolerance limits \parencite{zhang2007effect, paganetti2021adaptive}. To mitigate these effects, many centers employ daily image guidance and adaptive workflows \parencite{paganetti2021adaptive, albertini2020online}.

Most modern proton therapy facilities are equipped with cone beam computed tomography (CBCT) systems for daily setup verification \parencite{hua2017robotic}. CBCT provides three-dimensional anatomical information at the treatment position, making it an attractive modality for dose calculation in adaptive therapy workflows. However, direct dose computation on CBCT is challenged by scatter, beam hardening, motion artefacts, and limited soft tissue contrast relative to planning CT \parencite{liu2023review,giacometti2020review}.

A core difficulty is mapping CT numbers (Hounsfield units, HU) to stopping-power ratios for accurate proton range calculations \parencite{landry2019comparing}. CBCT often exhibits systematic HU biases of \~50–100 HU vs fan-beam CT used for treatment planning, with spatial non-uniformities that can translate to $>$5 mm range uncertainty in heterogeneous regions \parencite{peters2023consensus}. These effects degrade dose accuracy and motivate correction strategies.

Traditional approaches to CBCT-based dose calculation employ multi-step correction workflows that attempt to address image quality limitations through various methodologies \parencite{liu2023review,giacometti2020review}. Scatter correction algorithms utilize physical models or measurement-based approaches to estimate and subtract scatter contributions from CBCT projections \parencite{trapp2022empirical}. In contrast, synthetic CT generation methods employ image registration techniques to deform planning CT images to match daily CBCT anatomy, creating hybrid images with improved CT number accuracy \parencite{landry2015investigating}. Alternative approaches include histogram matching \parencite{arai2017feasibility}, intensity correction \parencite{kurz2015comparing}, and density override \parencite{dunlop2015comparison} methods that aim to standardize CBCT images for dose calculation applications.

Despite technological advances, these correction methods introduce additional computational complexity, potential error propagation, and workflow inefficiencies that limit their practical implementation in time-constrained adaptive therapy protocols \parencite{shen2020introduction}. The multi-step nature of traditional approaches increases the overall treatment planning time and introduces multiple potential failure points that require quality assurance verification \parencite{liu2023review}. Furthermore, the accuracy of correction methods often depends on the quality of the original CBCT images and the magnitude of anatomical changes, leading to variable performances across different clinical scenarios \parencite{giacometti2020review}.

The emergence of deep learning methodologies in medical physics has opened new possibilities for addressing the CBCT dose calculation challenge through end-to-end learning approaches \parencite{cui2020introduction,wang2021review}. Recent studies have explored various neural network architectures, including convolutional networks, Long Short-Term Memory (LSTM) \parencite{hochreiter1997long}, and Transformer \parencite{vaswani2017attention} approaches for dose prediction applications \parencite{neishabouri2021long,pastor2022millisecond}.

Sequence modelling approaches, particularly extended Long Short-Term Memory (xLSTM) architectures \parencite{beck2024xlstm}, have shown exceptional promise for medical imaging applications. The xLSTM architecture offers enhanced memory mechanisms, improved computational efficiency, and superior performance compared to traditional LSTM implementations. Recent applications in MRI-based dose calculation have demonstrated close to Monte Carlo accuracy with substantial speed improvements, establishing the feasibility of sequence-based approaches for radiotherapy planning \parencite{li2025proof}.

Building upon these advances, this study investigates the adaptation of xLSTM-based neural network architectures for direct CBCT-based proton dose calculations. The approach leverages the sequence modelling capabilities of xLSTM to capture complex spatial relationships in beam-wise dose deposition patterns while eliminating the need for traditional CBCT correction workflows. The methodology enables direct transformation from CBCT images to accurate dose distributions, with potential advantages in adaptive proton therapy workflows through improved computational efficiency and enhanced clinical practicality.

The primary objectives of this study were to validate a deep learning approach against Monte Carlo reference calculations using clinical patient data with realistic anatomical variations, to develop a deep learning-based dose calculation engine capable of accurate proton dose prediction directly from CBCT images, and to characterize model performance across different beam and anatomical configurations. Secondary objectives included assessment of dose-volume histogram preservation for critical structures and demonstration of computational efficiency suitable for integration into clinical treatment planning workflows.

\section{Methods}
\subsection{Dataset and Patient Cohort}
This retrospective study utilized paired CT-CBCT images from 45 head-and-neck cancer patients treated with intensity-modulated proton therapy at our institution between 2021 and 2023. The dataset comprised 77 paired CT-CBCT images, with each patient contributing 1-4 image pairs. All patient data were anonymized for research purposes following institutional review board guidelines.

Patient imaging was performed using standardized clinical protocols to ensure consistency across the dataset. Planning CT images were acquired using a Siemens Sensation Open scanner with 120 kV and reconstructed at 1 mm × 1 mm × 2 mm voxel size with a soft tissue kernel. Daily CBCT images were obtained using ProBeam® Proton Therapy System with a standard head-and-neck protocol (Varian Medical Systems, Palo Alto, CA, USA), reconstructed at 0.56 mm × 0.56 mm × 2 mm resolution and subsequently resampled to match planning CT resolution.

Image registration was performed using a two-step approach to ensure accurate anatomical alignment between planning CT and corresponding CBCT images. Initial rigid registration was performed using the ANTsPy \parencite{tustison2021antsx} library with mutual information-based algorithms to establish global alignment. Subsequently, deformable registration was applied using an internally developed deformable registration tool \parencite{li2024continuous} to account for soft tissue deformations. Registration accuracy was verified through comprehensive visual inspection to ensure adequate anatomical alignment quality.

For model development, the dataset was systematically divided into training, validation, and independent test cohorts following standard machine learning practices. The training dataset included 33 patients (55 image pairs) for model parameter optimization. The validation dataset comprised 7 patients (12 image pairs) for hyperparameter tuning, early stopping criteria, and model selection during the development process. Critically, an independent test dataset comprised 5 additional patients who were completely held out from all aspects of model development, including training, validation, and hyperparameter selection. These 5 patients represent truly unseen data and were used exclusively for final clinical performance evaluation. The independent test patients underwent patient-specific fine-tuning (using only their planning CT data) followed by comprehensive validation on their treatment CBCT images to assess clinical performance under realistic adaptive therapy scenarios.

\subsection{Beam Configuration Sampling and Data Augmentation}
Comprehensive beam sampling was performed based on the PSI Gantry2 proton therapy setup \parencite{pedroni2011pencil} to generate a diverse training dataset. For each CT-CBCT image pair, 1500 proton beams were systematically sampled using randomized parameters centered on the image isocenter, with stochastic variations to reflect clinical scenarios.

Beam sampling employed a probabilistic approach with the following parameters: gantry angles were randomly sampled from -30° to 180°, and couch angles from -180° to 180°. Nozzle extraction distances were sampled from 1 to 27 cm, representing the full range of air gaps available with the PSI Gantry2 configuration and enabling simulation of various beam delivery scenarios from close-proximity to extended-distance treatments. Field centers were positioned at the image volume center with random spatial offsets to account for setup variations and target positioning uncertainties.

Each beam configuration included a single pencil beam spot with random 2D lateral offsets ranging from -10 to +10 cm to sample different beam positions within the field. Beam weights were set to 1000 monitor units (MU), corresponding to approximately $10^4$ protons per MU, ensuring consistent total particle numbers for dose calculations across all configurations. Energy selection utilized a probabilistic distribution derived from the beam energies used in the 40 clinical head-and-neck treatment plans from the training and validation datasets. All beam energy values from these plans were collected and fitted to create a sampling distribution that accurately reflects institutional clinical practice. During training data generation, beam energies for randomly sampled configurations were drawn from this fitted distribution, ensuring the training dataset energy spectrum matched actual clinical energy utilization patterns.

Beam's-eye-view (BEV) patch extraction represents a critical methodological component that transforms the 3D dose calculation problem into a sequence modeling task suitable for xLSTM processing. For each beam configuration, the CT/CBCT volume was geometrically transformed and resampled to create a beam's-eye-view perspective aligned with the proton beam direction. The extraction process involved: (1) coordinate system transformation to align the image volume with the beam axis, (2) definition of the extraction volume centered on the beam path, and (3) systematic resampling to create standardized patch dimensions. Each BEV volume encompassed 24 × 24 × 255 voxels at 2 mm isotropic resolution, where the 24×24 transverse dimensions correspond to a 4.8×4.8 cm field-of-view and the 255-voxel depth dimension (51 cm total depth) provided adequate coverage for full-range proton penetration in head-and-neck anatomy. The depth dimension was designed to encompass the complete proton range for clinical energies while maintaining computational tractability for neural network training and inference operations.

\subsection{Monte Carlo Simulation and Ground Truth Generation}
Ground truth dose distributions were generated using the FRED Monte Carlo simulation platform \parencite{schiavi2017fred}, which has been validated for proton therapy applications \parencite{gajewski2021commissioning}. Simulation parameters were optimized to ensure statistical accuracy while maintaining computational feasibility for large-scale dataset requirements.

Each beam simulation utilized 1 million primary proton histories to achieve statistical uncertainties below 1\% in high-dose regions ($>$50\% of maximum dose) and below 2\% in intermediate-dose regions (10-50\% of maximum dose). Dose grid resolution was maintained at 2 mm isotropic spacing to match BEV patch dimensions and provide adequate spatial resolution for capturing typical lateral dose gradients in proton therapy.

Output dose matrices were normalized to absolute dose per beam and subsequently scaled to clinical prescription levels for training purposes. Parallel computing infrastructure utilizing NVIDIA RTX 4090 GPUs enabled efficient Monte Carlo calculations, with average simulation times of 2 seconds per beam configuration.

\begin{figure}[htbp]
    \centering
    \includegraphics[width=1.0\textwidth]{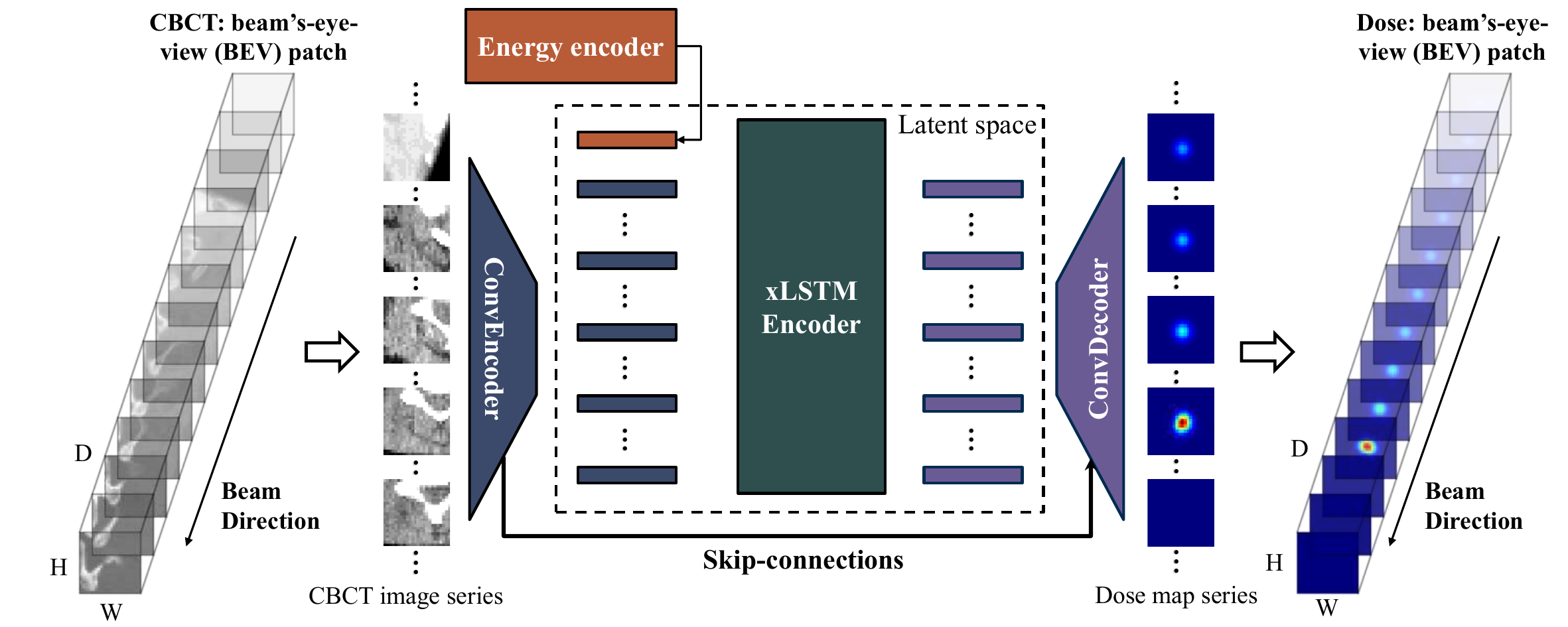} % 第一张图
    \caption{CBCT-NN method overview and processing pipeline.}
    \label{fig:fig1}
\end{figure}

\subsection{Neural Network Architecture and Implementation}
The CBCT-NN architecture implemented an encoder-decoder framework enhanced with xLSTM-based sequence modeling specifically designed for beam-wise proton dose prediction (Figure~\ref{fig:fig1}). The architecture follows a modular design that processes beam's-eye-view patches through hierarchical feature extraction, sequence modeling, and dose reconstruction stages.

The encoder component (ConvEncoder) utilized a hierarchical 2D convolutional neural network structure with three encoding blocks to process beam's-eye-view patches slice by slice. The first convolutional layer employed 64 feature channels with 5×5 kernels, followed by group normalization (16 groups) and SiLU activation functions. The second convolutional layer maintained 64 channels with identical kernel configuration, while the third layer projected features to the desired embedding dimension. Max pooling operations with 2×2 kernels and stride 2 were applied after the first two convolutional layers to progressively reduce spatial dimensions while preserving essential spatial information.

Spatial positional encoding was implemented using learnable embeddings that preserved sequential relationships across the depth dimension of BEV patches. These embeddings were crucial for maintaining spatial coherence during sequence modeling operations and enabling accurate dose gradient reconstruction in subsequent processing stages.

An innovative energy token encoding system represented a key architectural advancement, enabling the model to account for beam energy variations during dose prediction. In this, beam energy information was processed through an embedding layer for discrete energy categories, generating energy-specific representations that were concatenated with spatial features at the beginning of the sequence. This approach ensured that energy-dependent physics effects, including range modulation and lateral scattering variations, were appropriately captured throughout the dose prediction process.

The xLSTM sequence modeling component formed the core innovation of architecture, leveraging enhanced memory mechanisms to capture complex spatial dependencies in dose deposition patterns \parencite{beck2024xlstm}. The xLSTM encoder utilized a block stack configuration incorporating both matrix LSTM (mLSTM) and scalar LSTM (sLSTM) variants. The mLSTM blocks employed 4-head attention mechanisms with 1D convolutional kernels (size 4) and QKV projection block sizes of 4, while sLSTM blocks featured CUDA-optimized backends with 4-head attention and power-law block-dependent bias initialization. The configuration included 2 blocks total with sLSTM positioned at the second block, dropout rate of 0.2, and feedforward networks with 1.3× projection factors and GELU activation functions.

The decoder architecture (ConvDecoder) implemented symmetric expansion paths that progressively reconstructed full-resolution dose distributions from encoded sequence features. The decoder featured three main processing stages with skip connections from the encoder's convolutional layers. Each stage employed 5×5 convolutional kernels with group normalization and SiLU activation functions, followed by nearest-neighbor upsampling (factor 2) to restore spatial resolution.

Skip connections between corresponding encoder and decoder levels preserved fine-grained spatial information by concatenating encoder feature maps with decoder features at matching spatial resolutions. The decoder processed the sequence output by first removing the energy token through slicing operations and reshaping the remaining tokens to match the spatial dimensions, then applying the hierarchical reconstruction process to produce the complete 3D dose distribution.

The final output layer consisted of a single 5×5 convolutional filter that generated the predicted dose distribution for each BEV slice, with the complete output reshaped to match the original beam geometry for clinical dose calculation applications.

\subsection{Training Protocol and Optimization Strategy}
Model training employed a multi-stage approach designed to maximize performance while ensuring stable convergence and generalization to unseen data. The training protocol incorporated population-based pre-training followed by patient-specific fine-tuning to balance between broad applicability and individualized accuracy.

The initial pre-training phase utilized the dataset of 82,500 paired BEV patches (55 image pairs × 1500 beams). The LAMB optimizer was employed with an initial learning rate of 1e-3, which was reduced by a factor of 0.5 when validation loss plateaued for more than 10 and 20 epochs. The loss function utilized mean absolute error (MAE) between predicted and reference dose distributions to ensure dose calculation accuracy. This simplified loss formulation focused model optimization on fundamental dose prediction accuracy while maintaining training stability.

\subsection{Validation and Performance Metrics}
Comprehensive validation was performed using an independent test dataset of 5 patients (separate from both the 33-patient training and 7-patient validation cohorts), each with treatment plans optimized on the planning CT. For each patient, two distinct imaging pairs were utilized: (1) a planning-phase CBCT-CT pair for fine-tuning, where the CBCT was acquired during the first treatment fraction (temporally closest to planning CT acquisition), and (2) a late-treatment CBCT-CT pair for testing. The test data pair consisted of the CBCT and control CT acquired on the same day during the treatment fraction closest to the end of the treatment course, ensuring evaluation under conditions of maximum inter-fractional anatomical change.

Patient-specific fine-tuning was implemented to simulate realistic adaptive workflow scenarios and represents a key methodological innovation of this approach. For each test patient, the pre-trained population model was fine-tuned using their planning CT data to adapt the model to patient-specific anatomical features, tissue compositions, and geometric characteristics. This fine-tuning process involved generating approximately 1500 beam configurations from the planning CT using the same sampling methodology described above, followed by training for 100 epochs using a reduced learning rate of 2.5e-4 to prevent overfitting while optimizing patient-specific accuracy. The approach mimics a realistic clinical implementation scenario where the initial planning CT scan would be used to calibrate the neural network model for each individual patient prior to treatment delivery, enabling subsequent rapid dose calculations on daily CBCT images throughout the treatment course.

Image pairing and registration procedures were carefully implemented to ensure accurate validation conditions. Planning CT and treatment CBCT images were paired based on temporal proximity (CBCT acquired within 1-2 weeks of planning CT) and anatomical consistency verified through visual inspection by experienced medical physicists. A two-step registration process was performed between each CT-CBCT pair: initial rigid registration using mutual information-based algorithms, followed by deformable registration using an internally developed deformable registration tool to account for soft tissue deformations and anatomical changes between imaging sessions. Registration quality was assessed by examining anatomical landmark correspondence for both bony structures and soft tissue boundaries after the complete registration process.

Contour propagation for dose-volume histogram analysis was performed using the established registration transformations. Target volumes and organ-at-risk contours originally defined on the planning CT were propagated to the corresponding CBCT images using the registration parameters. Contour accuracy was verified through visual inspection and manual adjustment where necessary to ensure anatomical correspondence. This approach enabled consistent DVH comparisons between Monte Carlo calculations on repeated CT images and CBCT-NN predictions on corresponding CBCT images while maintaining identical geometric conditions for evaluation.

The validation protocol involved recalculating initial treatment plans on both repeated CT (using FRED Monte Carlo simulation) and corresponding CBCT (using CBCT-NN) to enable direct comparison under identical geometric conditions. This approach eliminated potential confounding factors related to plan optimization differences and focused evaluation on fundamental dose calculation accuracy.

Gamma analysis served as the primary validation metric, implemented using standard clinical criteria of 2mm/2\% and 2mm/3\% with global normalization and a 10\% low-dose threshold. Mean percentage dose error (MPDE) analysis was performed at various dose threshold levels to assess prediction accuracy across different dose regions. MPDE was calculated as the mean absolute difference between NN predicted and reference MC doses, normalized by the prescription dose, for voxels receiving doses above specified thresholds (5\%, 10\%, 50\%, and 90\% of maximum dose). This analysis provided complementary and more sensitive information to the gamma evaluation by quantifying dose accuracy in clinically relevant dose regions.

Dose-volume histogram analysis provided clinically relevant validation through comparison of key parameters for target volumes and organs at risk. Primary endpoints included clinical target volume (CTV) coverage (V95\%), parotid gland mean doses for xerostomia assessment, and spinal cord maximum doses.

Computational performance evaluation measured inference times for both single-beam calculations and complete treatment plans to assess clinical implementation feasibility. Timing measurements were conducted using clinical-grade hardware configurations (NVIDIA RTX 4090 GPUs with 24 GB memory) representative of modern treatment planning environments, with multiple repeated measurements to ensure reliability and reproducibility of performance assessments.

\section{Results}
\subsection{Overall Model Performance and Dose Calculation Accuracy}
The CBCT-NN model demonstrated good performance in direct dose calculation from CBCT images, achieving dose calculation accuracy suitable for clinical implementation across all validation metrics (Table~\ref{tab:gamma_results}). Gamma analysis results using 2mm/2\% criteria revealed a mean pass rate of 95.1 ± 2.7\% across all test cases. These results are comparable to previous neural network-based proton dose prediction studies that employed similar gamma analysis metrics on planning CT images \parencite{neishabouri2021long,pastor2022millisecond}, demonstrating that direct CBCT-based prediction can achieve similar accuracy to established CT-based neural network approaches. Individual patient results showed consistent performance, with pass rates ranging from 91.0\% to 97.5\%.

\setlength{\tabcolsep}{1.5pt} % 默认6pt
\begin{table}[]
\caption{Comprehensive model performance metrics and clinical validation results for CBCT-NN across 5 test patients. Dose percentages are reported relative to the prescribed dose.}
\centering
\begin{tabular}{c|c|c|cc|cc|cc}
\toprule
\multirow{2}{*}{Patient} & Gamma (2mm/2\%) & Gamma (2mm/3\%) & \multicolumn{2}{c|}{CTV V95 (\%)} & \multicolumn{2}{c|}{Parotid $D_{mean}$ (\%)} & \multicolumn{2}{c}{Spinal Cord $D_{max}$ (\%)} \\ \cline{2-9} 
 & CT-MC vs CBCT-NN & CT-MC vs CBCT-NN & CT-MC & CBCT-NN & CT-MC & CBCT-NN & \multicolumn{1}{l}{CT-MC} & \multicolumn{1}{l}{CBCT-NN} \\ \hline
1 & 91.0 & 92.2 & 95.2 & 92.6 & 62.8 & 61.3 & 44.2 & 44.3 \\
2 & 97.5 & 97.8 & 99.5 & 99.6 & -- & -- & -- & -- \\
3 & 97.1 & 98.0 & 97.2 & 96.8 & 57.5 & 58.6 & 67.0 & 65.2 \\
4 & 96.0 & 97.2 & 99.2 & 98.8 & 39.7 & 40.1 & 48.2 & 46.2 \\
5 & 94.2 & 95.0 & 98.5 & 98.7 & 47.3 & 45.3 & 11.8 & 15.4 \\ \bottomrule
\end{tabular}
\label{tab:gamma_results}
\end{table}

\setlength{\tabcolsep}{6pt} % 默认6pt
\begin{table}[]
  \caption{%
    Mean Percentage Dose Error (MPDE, \%) by dose region.
    MPDE is computed as the mean absolute percentage difference.
  }
\centering
\begin{tabular}{c|ccc|c}
\toprule
Patient & \textbf{$>90\%\,D_{\max}$} & \textbf{$50$--$90\%\,D_{\max}$} & \textbf{$10$--$50\%\,D_{\max}$} & Overall (body) \\ \hline
1 & 2.46 & 4.31 & 8.39 & 7.98 \\
2 & 2.13 & 3.87 & 4.14 & 4.39 \\
3 & 1.12 & 2.02 & 5.42 & 4.60 \\
4 & 2.33 & 2.72 & 4.67 & 4.63 \\
5 & 4.97 & 3.63 & 8.35 & 7.88 \\ \bottomrule
\end{tabular}
\label{tab:mpde_results}
\end{table}

\subsection{Mean Percentage Dose Error Analysis}
\begin{figure}[htbp]
    \centering
    \includegraphics[width=0.9\textwidth]{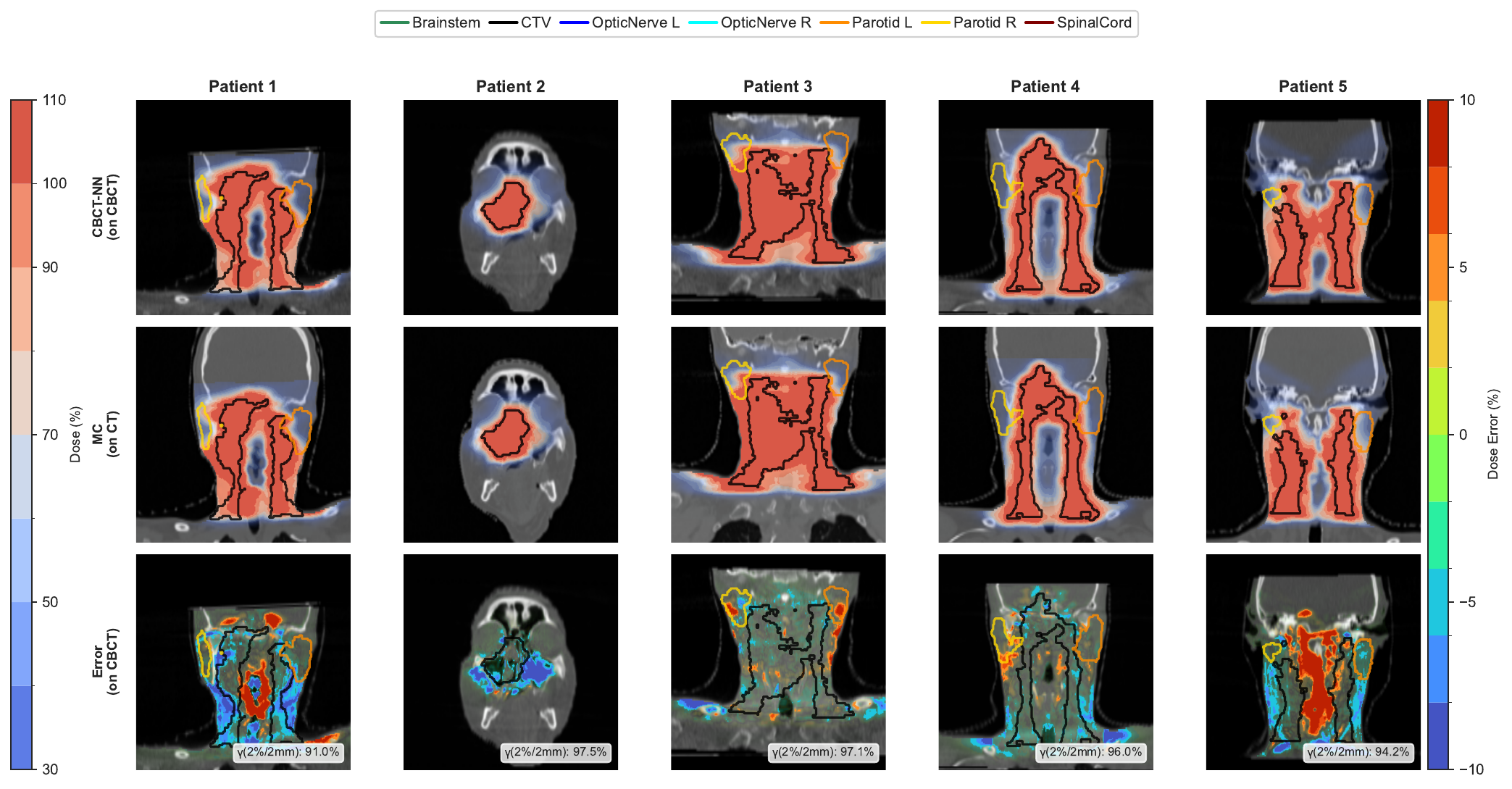}\\ % 第一张图
    \includegraphics[width=0.9\textwidth]{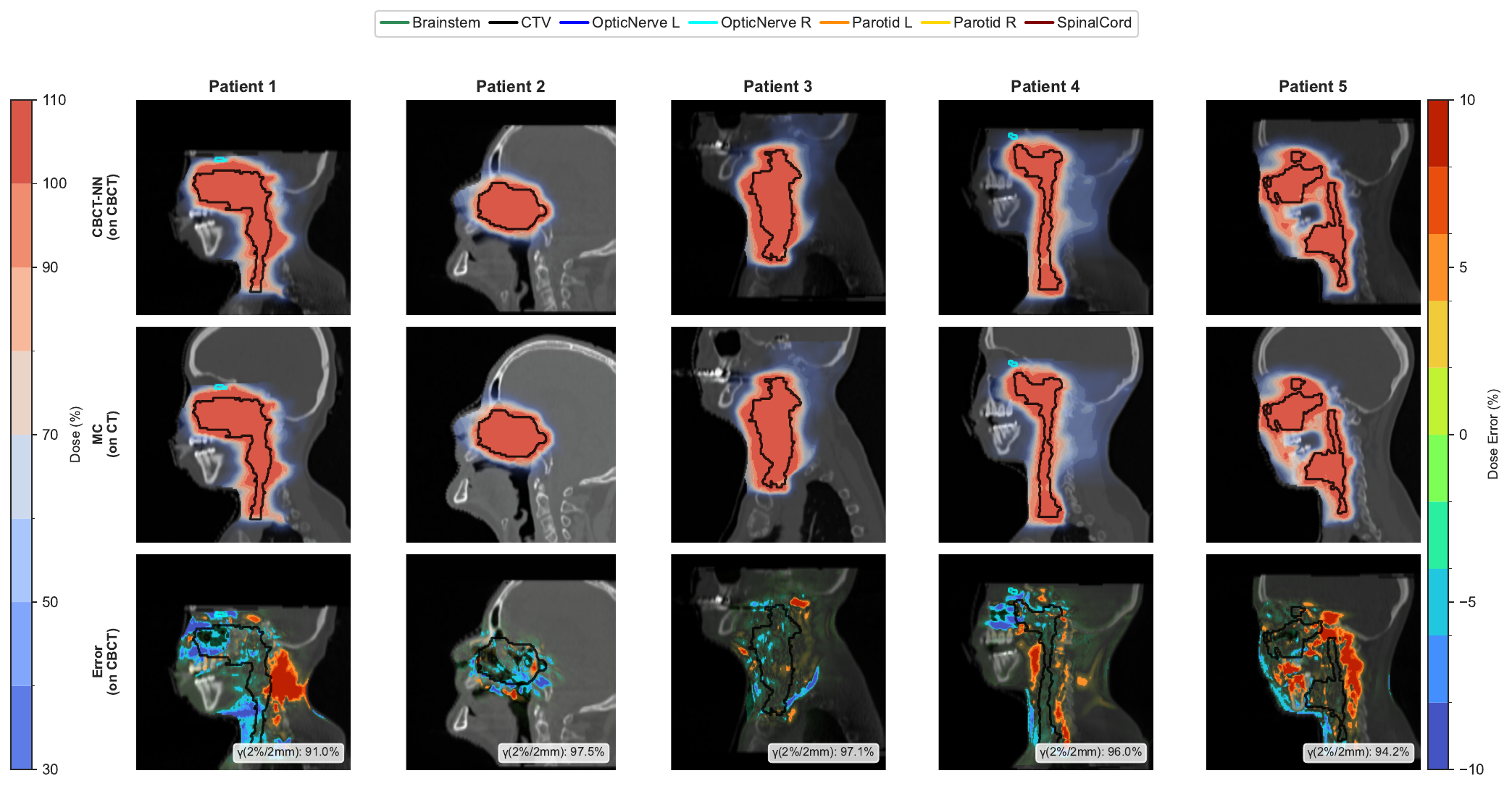}        % 第二张图
    \caption{Spatial dose distribution analysis and error visualization comparing CBCT-NN predictions with Monte Carlo reference calculations across five patient cases with varying anatomical complexity.}
    \label{fig:fig2}
\end{figure}

Mean percentage dose error analysis (calculated as mean absolute differences) provided detailed assessment of prediction accuracy across clinically relevant dose regions. For high-dose regions ($>$90\% of maximum dose), the mean MPDE was 2.6 ± 1.4\%. Intermediate-dose regions (50-90\% maximum dose) showed mean MPDE of 3.3 ± 0.9\%, while lower-dose regions (10-50\% maximum dose) achieved 6.2 ± 2.0\%. The overall body MPDE averaged 5.9 ± 1.9\%. Detailed per-patient values are summarized in Table~\ref{tab:mpde_results}. Higher accuracy was observed in high-dose regions compared to lower-dose regions. The spatial distribution of dose errors is illustrated through detailed dose maps and error visualizations (Figure~\ref{fig:fig2}).

\subsection{Dose-Volume Histogram Analysis and Clinical Relevance}
\begin{figure}[htbp]
    \centering
    \includegraphics[width=1.0\textwidth]{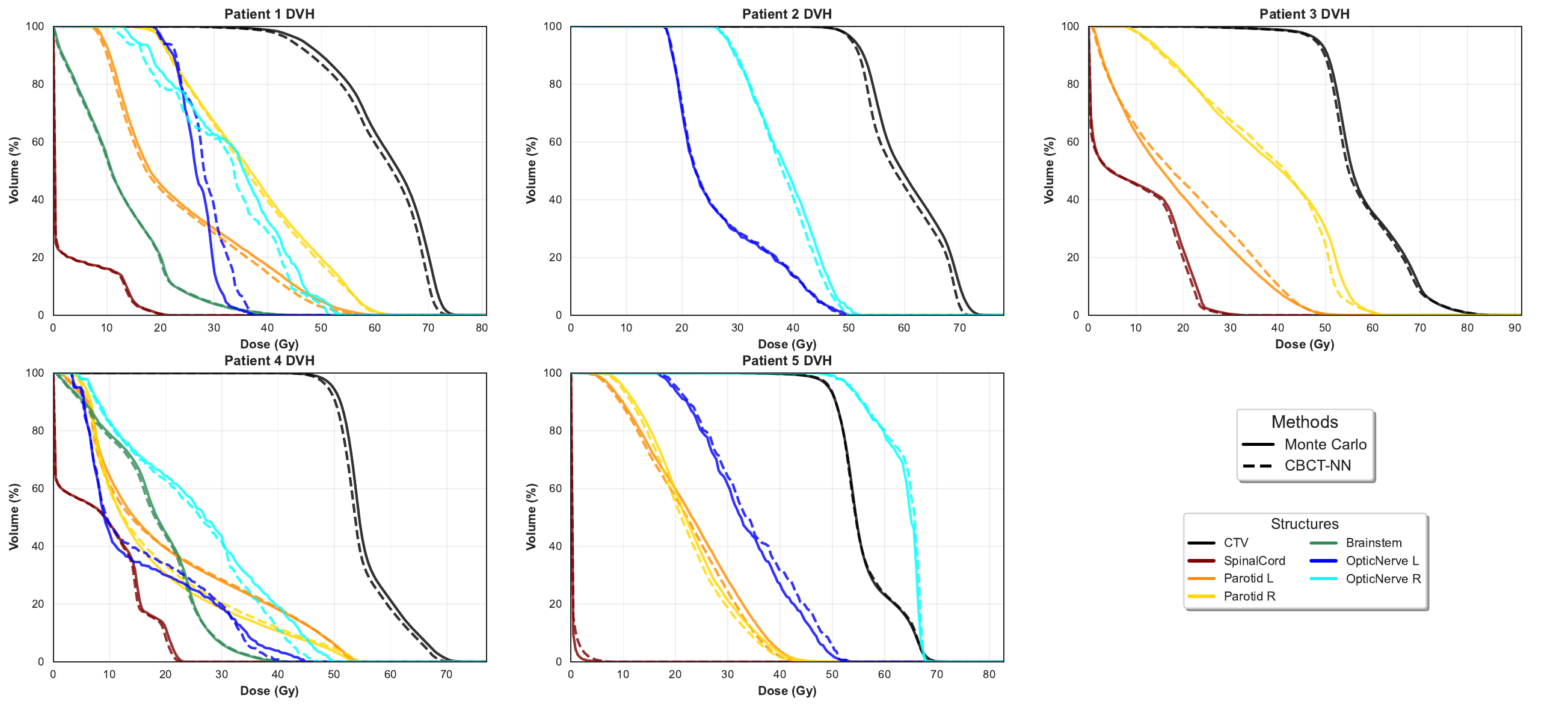} % 第一张图
    \caption{Dose-volume histogram comparison for target volumes and critical organs comprehensive DVH analysis comparing CBCT-NN predictions (dashed lines) with Monte Carlo reference calculations (solid lines) across multiple patient cases.}
    \label{fig:fig3}
\end{figure}

DVH analysis demonstrated good preservation of clinical parameters (Figure~\ref{fig:fig3}). Target volume analysis revealed minimal deviations between CBCT-NN predictions and Monte Carlo reference calculations. Clinical target volume (CTV) V95\% showed mean differences of -0.6 ± 1.1\%, indicating a slight conservative bias (Table~\ref{tab:gamma_results}).

OAR analysis showed good performance for critical structures. Parotid gland mean dose predictions had mean differences of -0.5 ± 1.5\% compared to Monte Carlo calculations. Spinal cord maximum dose assessment revealed mean differences of 0.1 ± 2.5\%. Close agreement between predicted and reference DVH curves for all critical structures is demonstrated across multiple patient cases (Figure~\ref{fig:fig3}).

\subsection{Computational Performance and Treatment Planning Efficiency}
Computational performance analysis demonstrated timing characteristics suitable for clinical implementation using clinical-grade hardware (NVIDIA RTX 4090 GPUs with 24 GB memory).

Individual beam dose calculations required an average of 3 milliseconds per pencil beam. Complete treatment plan calculations for typical head-and-neck cases (40,000-50,000 pencil beams) were completed in 1-3 minutes. The neural network approach provides substantial computational advantages over traditional Monte Carlo methods, though specific comparisons depend on Monte Carlo implementation and required statistical accuracy.

\section{Discussion}
This study establishes a proof-of-concept for direct CBCT-based proton dose calculation using xLSTM neural networks, achieving accurate dose prediction without traditional correction workflows. The gamma pass rates of 95.1 ± 2.7\% using 2mm/2\% criteria are consistent with previous neural network-based proton dose prediction studies on planning CT images \parencite{neishabouri2021long,pastor2022millisecond}, demonstrating that our direct CBCT-based approach achieves comparable accuracy to established CT-based deep learning methods while eliminating the need for image correction workflows. The computational efficiency of under 3 minutes for complete treatment plans significantly enhances the practical utility of CBCT-based dose calculations.

The xLSTM-based architecture represents a significant methodological advancement over traditional sequence modeling for radiotherapy applications. Enhanced memory mechanisms enable capture of complex spatial dependencies while maintaining practical processing times \parencite{li2020behrt}. The energy token encoding system provides a novel approach to incorporating physical parameters, enabling dynamic adaptation to energy-dependent physics effects. The beam's-eye-view sequence modeling successfully transforms the complex 3D dose calculation problem into a tractable sequence prediction task while preserving essential spatial relationships.

Comparison with conventional approaches highlights several advantages. Traditional CBCT correction workflows require multiple processing steps including scatter correction, image registration, or synthetic CT generation, each introducing potential error sources and computational overhead. The direct prediction approach eliminates these intermediate steps while achieving comparable accuracy. Deep learning approaches using traditional convolutional architectures struggle with long-range spatial dependencies essential for accurate proton dose calculation, while transformer-based approaches face computational scalability challenges \parencite{fan2020data}. The xLSTM approach effectively balances modeling capability with computational efficiency.

The demonstrated accuracy and computational efficiency position this approach as particularly suitable for adaptive radiotherapy applications. The ability to rapidly calculate accurate doses directly from CBCT images eliminates time-consuming correction workflows, enabling practical implementation of online adaptive protocols \parencite{albertini2020online} where rapid dose assessment is critical for treatment adaptation decisions.

The patient-specific fine-tuning methodology presents both advantages and limitations. The primary advantage lies in adapting the model to individual patient characteristics using readily available planning CT data. However, this introduces an additional processing step requiring computational resources and time (approximately 30 minutes per patient). Future research could explore alternative personalization strategies, including few-shot learning approaches, to reduce computational overhead while maintaining patient-specific optimization benefits.

The comprehensive validation methodology provides robust evidence for clinical applicability. The focus on independent test data with realistic anatomical changes ensures reported performance reflects true clinical scenarios rather than optimistic laboratory conditions. The validation using repeated imaging sessions during actual treatment courses demonstrates model performance under conditions representative of adaptive therapy implementation.

Several considerations are important for broader clinical implementation. The current validation was performed on head-and-neck cancer patients, providing focused validation in anatomically consistent regions. Extension to other treatment sites including thoracic and abdominal regions represents a natural progression that will benefit from the established methodology while accounting for site-specific challenges such as respiratory motion and larger anatomical variations \parencite{alraddadi2021literature}. The demonstrated robustness in head-and-neck applications provides confidence for systematic expansion to additional anatomical sites.

The approach demonstrates robust performance within the validated domain, and future developments will enhance generalization capabilities. Multi-institutional implementation may benefit from system-specific adaptations to account for CBCT image quality variations across different imaging platforms, similar to current clinical practices for dose calculation calibration. The methodology's modular architecture facilitates such adaptations without requiring fundamental changes to the core approach.

Clinical implementation considerations include quality assurance protocols appropriate for neural network-based dose calculation systems. The demonstrated accuracy supports potential use as a primary dose calculation engine, while computational efficiency enables comprehensive secondary verification workflows. Integration with existing systems requires minimal infrastructure modifications, utilizing standard CBCT acquisition protocols and generating outputs compatible with commercial treatment planning systems. Uncertainty quantification represents an important area for continued development, particularly for adaptive therapy applications where prediction confidence assessment supports clinical decision-making \parencite{seoni2023application}.

Future research directions include extension to treatment planning optimization workflows where the neural network could enable rapid dose calculation for multiple energy configurations during plan optimization processes, integration with predictive modeling for proactive adaptive therapy, and development of uncertainty-aware networks that provide confidence estimates alongside dose predictions. Multi-modal integration with functional imaging could create comprehensive adaptive platforms optimizing treatment based on multiple information sources \parencite{talbot2020multimodal}. The computational advantages could enable sophisticated treatment planning approaches including robust optimization and real-time plan adaptation across various treatment modalities \parencite{unkelbach2018robust}.

The success of sequence modeling approaches suggests broader applications in radiotherapy workflows, including treatment planning optimization, quality assurance automation, and treatment outcome prediction. The methodological advances demonstrate the potential for neural network-based approaches to enhance multiple aspects of radiation therapy planning and delivery, contributing to the development of more efficient and accurate treatment workflows.

\section{Conclusion}
This study validates the clinical feasibility of direct proton dose calculation on CBCT images using xLSTM neural networks, achieving Monte Carlo-level accuracy (95.1 ± 2.7\% gamma pass rates using 2mm/2\% criteria) with rapid computational performance. The approach eliminates traditional CBCT correction workflows and demonstrates robust performance across varying beam configurations and anatomical presentations through validation on independent test patients with realistic inter-fractional anatomical changes. Methodological innovations including xLSTM architecture and energy token encoding establish new standards for neural network dose calculation in proton therapy. The combination of high accuracy and computational efficiency positions this approach as particularly suitable for adaptive radiotherapy applications where rapid, accurate dose calculation directly from treatment imaging is essential for clinical decision-making.

\ack{This research was supported by the project "Increased Precision for Personalized Cancer Treatment Delivery Utilizing 4D Adapted Proton Therapy (EPIC-4DAPT)," funded by the Swiss National Science Foundation (SNSF) under grant number 212855. Evangelia Choulilitsa has received funding from the European Union's Horizon 2020 Marie Skłodowska-Curie Actions under Grant Agreement No. 955956.}
% \funding{...}
% \roles{...}
% \data{...}
% \suppdata{...}

% \section*{References}
\printbibliography
\end{document}